\documentclass[journal,twoside,web]{ieeecolor}
\usepackage{generic}
\usepackage{cite}
\usepackage{amsmath,amssymb,amsfonts}
\usepackage{newfloat}
\usepackage{listings}
\usepackage{multirow}
\usepackage{booktabs}
\usepackage{color}
\usepackage{algorithmic}
\usepackage{graphicx}
\usepackage{algorithm,algorithmic}
\usepackage{hyperref}
\usepackage{bbding}

\let\oldAA\AA
\renewcommand{\AA}{\text{\normalfont\oldAA}}

\def \m {\mathbf{m}}
\def \h {\mathbf{h}}
\def \z {\mathbf{z}}
\def \w {\mathbf{w}}

\usepackage{textcomp}
\def \BibTeX{{\rm B\kern-.05em{\sc i\kern-.025em b}\kern-.08em
    T\kern-.1667em\lower.7ex\hbox{E}\kern-.125emX}}
\begin{document}
\title{ETDock: A Novel Equivariant Transformer for Protein-Ligand Docking}
\author{Yiqiang Yi, Xu Wan, Yatao Bian, Le Ou-Yang,\IEEEmembership{Member, IEEE}, and Peilin Zhao
\thanks{This work was supported  by the National Natural Science Foundation of China [62173235,61602309], Guangdong Basic and Applied Basic Research Foundation [2022A1515010146, 2019A1515011384], and the (Key) Project of Department of Education of Guangdong Province [No. 2022ZDZX1022]. Corresponding author: Le Ou-Yang (email: leouyang@szu.edu.cn) and Peilin Zhao (email: masonzhao@tencent.com) }
\thanks{Yiqiang Yi, Xu Wan and Le Ou-Yang are with the  College of Electronics and Information Engineering, Shenzhen University, Shenzhen, 518060, China. (e-mail:yiyiqiang2021@email.szu.edu.cn; wanxu2021@email.szu.edu.cn; leouyang@szu.edu.cn)}
\thanks{Yatao Bian and Peilin Zhao are with Tencent AI Lab, Shenzhen, China.(e-mail: yatao.bian@gmail.com,masonzhao@tencent.com)}
}
\maketitle

\begin{abstract}
Predicting the docking between proteins and ligands is a crucial and challenging task for drug discovery. However, traditional docking methods mainly rely on scoring functions, and deep learning-based docking approaches usually neglect the 3D spatial information of proteins and ligands, as well as the graph-level features of ligands, which limits their performance. To address these limitations, we propose an equivariant transformer neural network for protein-ligand docking pose prediction. Our approach involves the fusion of ligand graph-level features by feature processing, followed by the learning of ligand and protein representations using our proposed TAMformer module. Additionally, we employ an iterative optimization approach based on the predicted distance matrix to generate refined ligand poses. The experimental results on real datasets show that our model can achieve state-of-the-art performance.
\end{abstract}


\section{Introduction}
\label{sec:introduction}

    
In the past few decades, great efforts have been made to study the structure and function of proteins \cite{du2021trrosetta,jumper2021highly,jin2022antibody}. The structure of biological entities has a significant impact on their functions \cite{giri2020multipredgo,zhou2022tasser}. However, predicting the structure of biological entities faces great challenges both experimentally and computationally\cite{fowler2022accuracy}. The advent of AlphaFold gave a major boost to the field of structural biology, which in turn had a profound impact on drug discovery \cite{jumper2021highly,sadybekov2023computational,nussinov2023alphafold}. In drug discovery, screening ligands by wet experiments is costly and time-consuming \cite{liu2023efficient,10027686}. Predicting the binding affinities between proteins and ligands computationally can facilitate the screening of drugs \cite{ma2023predicting,hua2023mfr,li2021structure}. However, the value of the binding affinity alone does not fully explain the interaction between a protein and a ligand\cite{lu2023improving,chu2022hierarchical,yang2023geometric}. To investigate the mechanisms underlying protein-ligand interactions more comprehensively, we need to predict protein-ligand docking, which is much more challenging than predicting their binding affinity \cite{lutankbind,stark2022equibind,ganea2021independent,yang2022protein}.

Protein-ligand docking is a widely used computational approach  for predicting and analyzing the interactions between proteins and small molecule ligands \cite{masters2022deep, stark2022equibind, lutankbind}, which plays a critical role in drug discovery and virtual drug screening \cite{chatterjee2023improving}. Traditional docking methods employ physics-based scoring functions and search algorithms to explore the chemical space and accomplish the docking process. AutoDock Vina utilizes an efficient search algorithm and scoring function to explore the conformational space of ligands \cite{trott2010autodock}. SMINA incorporates optimized search algorithms and scoring functions to predict accurate binding modes \cite{koes2013lessons}. GNINA enhances SMINA by incorporating a learned 3D Convolutional Neural Network (3D CNN) for scoring\cite{mcnutt2021gnina,tran2015learning}. QVina-W is a blind docking software that builds upon the speed-optimized QuickVina 2 by incorporating advanced algorithms for efficient exploration of ligand binding modes\cite{hassan2017protein}. GLIDE is a docking method that combines initial rough positioning, torsionally flexible energy optimization, and Monte Carlo sampling to achieve accurate ligand docking\cite{friesner2004glide}. However, traditional docking methods heavily rely on exhaustive conformational sampling of ligands and proteins within the vast chemical space, which is known to be computationally demanding and time-consuming.

Recently, some deep learning methods-based docking methods have been proposed to learn the scoring functions more accurately, but they often suffer from slower inference speeds due to their sampling-based frameworks. Tankbind  reduces the burden of conformational sampling by predicting a protein-ligand distance matrix and uses an optimization algorithm to convert the distance matrix into docking poses. However, 
Tankbind fails to account for the 3D spatial information of the ligand and protein\cite{lutankbind}. Equibind is an equivariant model that directly predicts the coordinates of binding pose structures. It refines the ligand conformations using graph neural networks and aligns the refined ligands to the binding pocket using a keypoint alignment mechanism \cite{stark2022equibind}. But Equibind does not consider the graph-level information of ligand, which limits the accuracy of its predictions.

To overcome the above challenges, we propose a novel \textbf{E}quivariant \textbf{T}ransformer \textbf{Dock}ing algorithm, named \textbf{ETDock}, for predicting protein-ligand docking pose. Our ETDock model includes two main modules, i.e., a feature processing module which integrates the atomic-level and graph-level information and a TAMformer module which consists of three layers to extract multi-level information from proteins and ligands for docking prediction. In particular, in the feature processing module, we first employ graph isomorphism networks (GIN) and graph vector prediction (GVP) to learn features from ligands and proteins, respectively \cite{xu2018powerful,jing2020learning}. To integrate graph-level and atom-level features, we introduce functional-class fingerprints (FCFPs) and fuse them with the atom-level features learned by GIN \cite{rogers2010extended}. Then we employ learnable outer products to capture the interaction features between ligands and proteins. These features are then fed into the module that consists of the triangle layer, the attention layer, and the message layer, called TAMformer. In the TAMformer module, the first layer is a triangle layer, which is designed to capture the physical constraints of ligand-protein interactions. The triangle layer can effectively encode geometric and spatial information to enforce structural constraints during docking prediction. The second layer is an attention layer, which focuses on extract relevant features of ligands and proteins to help predicting the docking pose. By employing our attention mechanisms, the model can selectively attend to important regions and interactions, enhancing the accuracy of the prediction. The third layer is a message layer, which facilitates the interaction between scalar and vector information of ligands and proteins. This allows for effective information exchange and integration between different components of the molecules. Furthermore, our ETDock generates a distance matrix between the ligand and protein, as well as confidence scores for different pockets. As ETDock is a two-stage model, ETDock utilizes the generated distance matrix to optimize the ligand pose iteratively.

By applying our ETDock model on the PDBbind v2020 dataset to predict protein-ligand docking, we have observed that our model possesses a remarkable inductive bias that aids in achieving superior performance\cite{liu2015pdb}. According to  our experiment results, ETDock outperforms traditional docking methods and deep learning-based docking approaches in predicting ligand docking poses. In addition, we thoroughly validate the efficacy of the feature fusion, triangle layer, attention layer, message layer and equivariant vector components.

The main contributions of this paper are summarized as follows:
\vspace{0.1in}
\begin{itemize}
     \item  We design a feature fusion block to integrate the atom-level and graph-level features of ligands, helping to capture more comprehensive representations of ligands.
    \vspace{0.1pt}
    \item  We develop a message layer to extract scalar and equivariant vector information from ligand, protein, and ligand-protein pair, which enables mutual learning of information from these three views and enhances the representation of both ligand and protein.
    \vspace{0.1in}
    \item We introduce a novel equivariant transformer framework for protein-ligand docking prediction. This framework utilizes equivariant vectors to model 3D spatial information and integrates chemical features with 3D spatial information to predict the ligand pose.
    \vspace{0.1in}
    \item Experimental results on the PDBbind v2020 datasets demonstrate that our proposed ETDock model outperforms previous traditional methods and deep learning approaches. Specifically, ETDock achieves an RMSD 23.2\% for ligand poses below 2\AA \space and an RMSD 61.1\% for ligand poses below 5\AA.
\end{itemize}

\section{Method}
\begin{figure*}[t]
\centering
\includegraphics[width=0.95\linewidth]{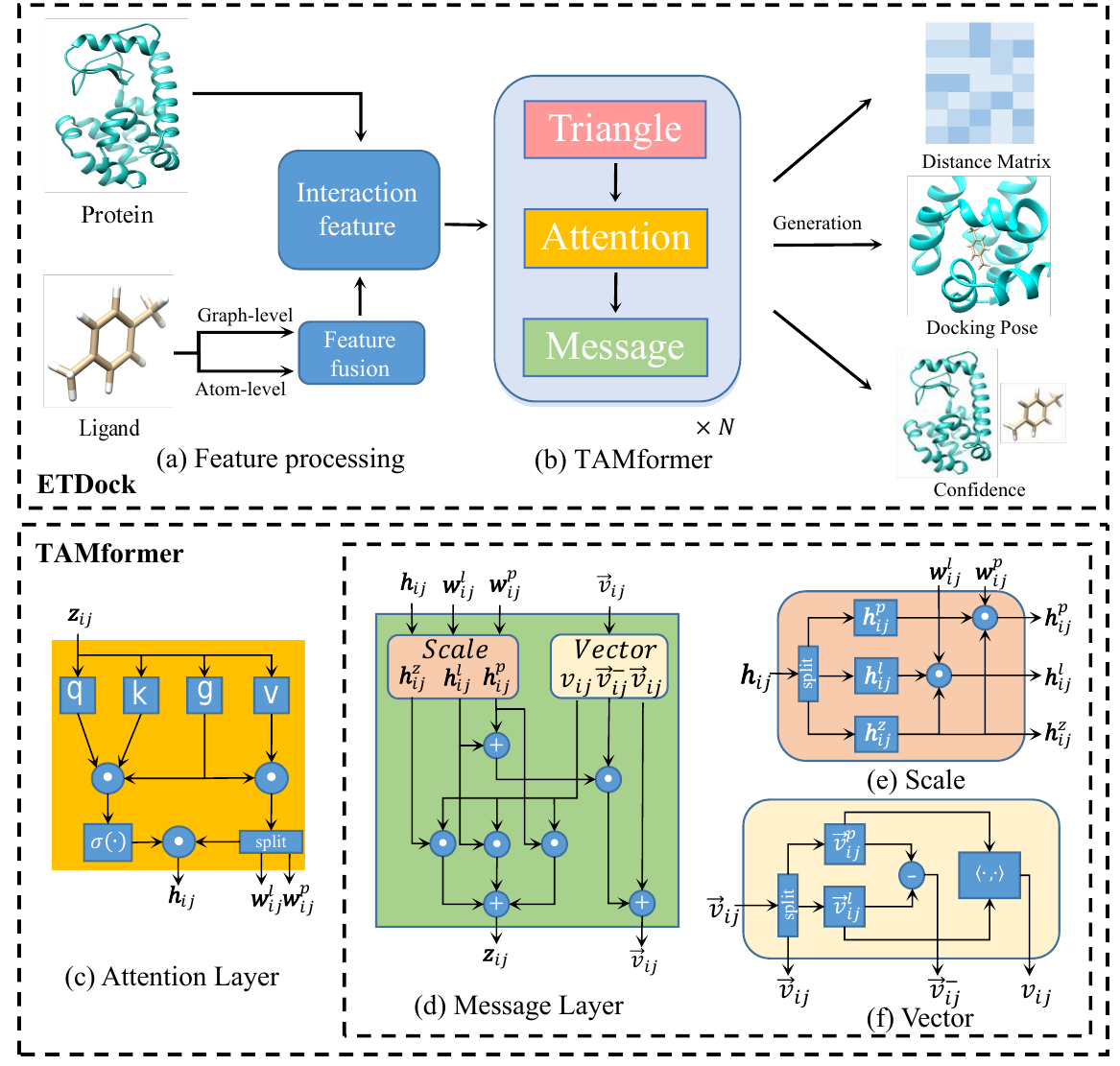} 
\caption{The overall framework of our ETDock model. (a) Feature Processing module: we fuse the atom-level and graph-level features of ligands and calculate the interactive features between ligands and proteins by performing outer product. (b) TAMformer module: consists of a triangle layer, an attention layer, and a message layer. Our model utilizes the triangle layer to capture the physical constraints between ligands and proteins. (c)  Attention layer: promotes the learning of ligand and protein features from interaction features. (d) Message layer: we perform message passing on the scalar and equivariant vector features of the ligand, protein, and protein-ligand interactive features to exchange information among them. (e) Scale module: we interact and update the scalar information of the ligand, protein, and protein-ligand  interactive features to exchange information among them. (f) Vector module: obtains relative features of the vector and compute scalar features through inner products of the vector features.
}
\label{fig:framework}
\end{figure*}

In this section, we present the details of our proposed \textbf{E}quivariant \textbf{T}ransformer \textbf{Dock}ing (\textbf{ETDock}) model for protein-ligand docking. The overall pipeline of our ETDock is illustrated in Fig. \ref{fig:framework}. Our model comprises three fundamental components: (1) Feature processing module which merges the atomic-level and graph-level features of ligands and learns the interactive characteristics between ligands and proteins. (2) TAMformer module that captures information from ligands and proteins by incorporating a triangle layer, an attention layer, and a message layer. (3) Ligand pose prediction generates a distance matrix for the protein-ligand complex and optimizes the docking pose based on this matrix iteratively.

\subsection{Feature Processing}
 The ligand is treated as a molecular graph $\mathcal{G}_l = (\mathcal{V}_l, \mathcal{E}_l)$ and the ligand node embeddings $M^l = [\m_1^l, ..., \m_i^l,...,\m_s^l]\in R^{d\times s}$ are learned by graph isomorphism network (GIN) \cite{xu2018powerful}, where $d$ is the embedding dimension, $s$ is the number of atoms in the ligand. Meanwhile, the learned atomic-level features $\m_i^{l}$ and the graph-level features $\m^{g} \in R^{1024\times 1}$ themselves are fused together to enable the model to better capture the complete information of the ligand. The protein $\mathcal{G}_p = (\mathcal{V}_p, \mathcal{E}_p)$ uses a K-nearest neighbor graph at the level of residues to represent and the protein node embeddings $M^p = [\m_1^p, ..., \m_j^p,...,\m_q^p]\in R^{d\times q}$ are learned by geometric vector perception (GVP) \cite{jing2020learning}, where $q$ is the number of atoms in the protein. Then using a learnable outer product to obtain the protein-ligand interaction features $\z_{i j}$.

 \subsubsection{Feature Fusion}
In previous methods, the focus was on learning atomic-level features while ignoring the graph-level features of ligands \cite{stark2022equibind,lutankbind}. This strategy limits the ability of the model to effectively learn the complete information of ligands. To overcome this limitation, we integrate the atomic-level and graph-level features (FCFPs) of ligands, allowing the model to obtain a more comprehensive understanding of the ligand properties \cite{deng2010multiple}. However, atomic-level and graph-level features are not in the same feature space, so transformation is carried out through the following equation:
\begin{equation}
\label{space_map}
\m^{g} = MLP(\m^{g}),
\end{equation}
where the $MLP(\cdot)$ is multilayer perceptron and maps the graph-level features to the atomic-level feature space to fuse in the same feature space.

Then, we fuse them in the same feature space through an attention mechanism, as shown in equation \eqref{feat_atten}-\eqref{feature_fusion}.
\begin{equation}
\label{feat_atten}
\w^{g}= softmax(\frac{(M^l)^\top \m^g}{\sqrt{d}}),
\end{equation}
\begin{equation}
\label{feature_fusion}
\m^{l}_{i}= \m^{l}_{i} + M^l\w^g,
\end{equation}
where $(M^l)^\top$ is the transpose of $M^l$, and  $\w^{g}$ is the attention of each atom for the whole ligand. This is inspired by previous works \cite{hua2022cpinformer,hua2023mfr}. We perform feature fusion by mapping the graph-level features to the atomic-level feature space and computing the attention of each atom to the entire ligand.

\subsubsection{Interaction Feature}
For the protein-ligand docking task, protein-ligand docking is imperative to model the interaction features between the protein and the ligand, and capture the complex interactions between them. Conventionally, outer product is employed to derive the interaction features between the ligand and the protein \cite{lutankbind}. However, this computation lacks the ability of parameter learning. To circumvent this limitation, we employ a learnable outer product operation as follows:
\begin{equation}
\label{learnable1}
\m_i^{l}= MLP(\m_i^{l}),
\end{equation}
\begin{equation}
\label{learnable2}
\m_j^{p}= MLP(\m_j^{p}),
\end{equation}
\begin{equation}
\label{interaction_feat}
\z_{ij}= \m^{l}_{i} \otimes \m^{p}_{j},
\end{equation}
where $\otimes$ is the outer product. We leverage MLP to introduce learnable parameters into the outer product operation, thereby enabling the outer product to learn parameters during computation. This facilitates the effective learning of deep interaction features by the model, which in turn can enhance the prediction accuracy of binding ligand poses.

\subsection{TAMformer: Triangle Attention Message  former}

\subsubsection{Triangle Layer} 
When a ligand binds a protein, the distances between different atoms can be cosidered as fixed, including $D^{l}_{i j},D^{p}_{i j}$ between the atoms within the protein and those in the ligand. Consequently, it is imperative to account for this constraint in the model. To achieve this, Tankbind utilizes the triangle layer to model this constraint\cite{lutankbind,jumper2021highly}, as follows:
\begin{equation}
\label{triangle}
\m^{tri}_{ij} = \sum_{k=1}^{q} D^{p}_{ik}\mathbf{t}_{kj}^{p} + \sum_{k^{'}=1}^{s}  \mathbf{t}_{ik^{'}}^{l} D^{l}_{k^{'}j} 
\end{equation}
\begin{equation}
\label{}
\z_{ij}= \z_{ij}+\m^{tri}_{ij}\odot gate(\z_{ij})
\end{equation}
where $\m^{tri}$ is the triangle message.
The physical constraint is integrated into the model via equation \eqref{triangle}, allowing for the effective modeling of the fixed distances between the atoms in the protein and the ligand during binding. where $\mathbf{t}_{ij}=MLP(\z_{ij})\odot gate(\z_{ij})$ and $gate(\z_{ij}) = \sigma(MLP(\z_{ij}))$.

\subsubsection{Attention Layer} 
Self-attention is commonly used to learn deep features of atoms in the molecular neighborhood \cite{vaswani2017attention,tholke2022torchmd}. We use self-attention to capture the deep interaction features between protein and ligand. Since the protein-ligand interaction features are obtained from individual protein and ligand through learnable outer product, they also contain information about protein and ligand themselves. However, traditional self-attention cannot simultaneously learn the individual information of protein and ligand from the protein-ligand interaction features. Therefore, we modify the self-attention by adding a gate to help the model learn the individual information of protein and ligand from the protein-ligand interaction features\cite{liao2022equiformer,morehead2021geometric}, as follows:

\begin{equation}
\label{gate}
    \mathbf{g}_{ij}= gate(\z_{ij}),
\end{equation}
\begin{equation}
\label{}
\mathbf{w}_{ij}^{z},\mathbf{w}_{ij}^{l},\mathbf{w}_{ij}^{p} = split(\mathbf{v}_{ij} \odot \mathbf{g}_{ij}),
\end{equation}
\begin{equation}
\label{}
\mathbf{a}_{ijk^{'}}^{z}= softmax(\mathbf{q}_{ij}^\top  \mathbf{k}_{ik^{'}} \odot \mathbf{g}_{ij}),
\end{equation}
\begin{equation}
\label{}
\h_{ij}^{z}= \sum_{k^{'}=1}^{s} \mathbf{a}_{ijk^{'}}^{z} \odot \mathbf{w}_{ij}^{z},
\end{equation}
where $split(\cdot)$ is used to divide the data into three parts. $\mathbf{q},\mathbf{k},\mathbf{v}$ are linear projections of the protein-ligand interaction feature $\z_{ij}$. $\mathbf{w}_{ij}^{z},\mathbf{w}_{ij}^{l},\mathbf{w}_{ij}^{p}$ is the value of protein-ligand interaction features, ligand, and protein, respectively. $\mathbf{a}_{ijk^{'}}^{z}$ is the self-attention weight of interaction feature.
$\h_{ij}^{z}$ is the protein-ligand interaction embedding. $\mathbf{g}_{ij}$ captures the information of ligand and protein from the value and captures the self-attention weight of the interaction features from the query and key. Finally, the interaction features output is obtained through $\mathbf{a}_{ijk^{'}}^{z}$ and $\mathbf{w}_{ij}^{z}$.

\subsubsection{Message Layer}
Previous methods have overlooked the incorporation of  chemical features and 3D spatial information from the ligand, protein, and ligand-protein pair perspectives\cite{lutankbind, stark2022equibind}. To overcome this, we introduce  equivariant vectors to learn the equivariant information of the ligands and protein, and update the vectors using the equivariant graph neural networks (EGNNs) paradigm \cite{satorras2021n}. Then, we enable interaction between invariant information and equivariant vectors. The scale block and vector block are used to process the invariant and equivariant vector before message passing.

The scale block apply the protein-ligand interaction feature to further interact with the ligand and protein, and utilize the values from the attention layer to further focus the ligand and protein, allowing the capturing of the ligand and protein information from the interaction feature by the following equations:
\begin{equation}
\label{}
\h_{ij}^{z},\h_{ij}^{l},\h_{ij}^{p}= split(\h_{ij}^{z}),
\end{equation}
\begin{equation}
\label{14}
\h_{ij}^{l} = \h_{ij}^{l}  \odot \h_{ij}^{z} \odot \mathbf{w}_{ij}^{l},
\end{equation}
\begin{equation}
\label{15}
\h_{ij}^{p} = \h_{ij}^{p}  \odot \h_{ij}^{z} \odot \textbf{w}_{ij}^{p},
\end{equation}
where $\h_{ij}^{z},\h_{ij}^{l},\h_{ij}^{p}$ is the embedding of protein-ligand interaction features, ligand, and protein, respectively. 

The vector block separates the protein-ligand interaction vector into ligand and protein vector features. The vector block learns the relative vector information between ligand and protein, as well as scalar features learned from the ligand and protein vector by the inner product.
\begin{equation}
\label{}
\overrightarrow{v}_{ij},\overrightarrow{v}_{ij}^{l},\overrightarrow{v}_{ij}^{p} = split(\overrightarrow{v}_{ij}),
\end{equation}
\begin{equation}
\label{}
\overrightarrow{v}_{ij}^{-} = \overrightarrow{v}_{ij}^{l} - \overrightarrow{v}_{ij}^{p},
\end{equation}
\begin{equation}
\label{}
v_{ij} = \left \langle  \overrightarrow{v}_{ij}^{l}, \overrightarrow{v}_{ij}^{p} \right \rangle,
\end{equation}
where $\overrightarrow{v}_{ij}, \overrightarrow{v}_{ij}^{l}, \overrightarrow{v}_{ij}^{p}$ are the vectors of interaction feature,  ligand feature and protein feature. We initialize $\overrightarrow{v}_{ij} = \overrightarrow{v}_i^{l} \otimes \overrightarrow{v}_j^{p}$, $\overrightarrow{v}_i^{l}$ by utilizing coordinates generated by the RDKit and initialize $\overrightarrow{v}_j^{p}$ using true coordinates \cite{landrum2013rdkit}.
$\overrightarrow{v}_{ij}^{-}$ is the relative vector information between ligand and protein. $\left \langle \cdot,\cdot \right \rangle$ is the inner prodcut. $v_{ij}$ is the scale feature from ligand and protein vector feature by scalar product.

In the message passing, we propagate information among protein-ligand pair, ligand, and protein. Prior to this step, we employ scale and vector block to process invariant and equivariant vector information. The invariant and equivariant vector information are  used to guide the learning of protein-ligand pair, ligand, and protein features through message passing. This way, the invariant and equivariant vector information serve as a means of communication and coordination between different features:
\begin{equation}
\label{19}
\begin{split}
    \h_{ij}^{z} = v_{ij}\h_{ij}^{z}, \\
    \h_{ij}^{l} = v_{ij}\h_{ij}^{l}, \\ 
    \h_{ij}^{p} = v_{ij}\h_{ij}^{p},
\end{split}
\end{equation}
\begin{equation}
\label{20}
\z_{ij} = \h_{ij}^{z} + \h_{ij}^{l} + \h_{ij}^{p}.
\end{equation}

After the message passing using equation \eqref{19}-\eqref{20}, we update the vectors using equation \eqref{21}, which is inspired by EGNNs, to maintain the equivariance of the learned features and make them more physically meaningful:
\begin{equation}
\label{21}
\overrightarrow{v}_{ij} = \overrightarrow{v}_{ij} + \psi (\h_{ij}^{l} + \h_{ij}^{p}) \overrightarrow{v}_{ij}^{-},
\end{equation}
where $\psi(\cdot)$ is the mean operation over the feature dimension. we incorporate invariant information of ligand and protein into the update of the protein-ligand interaction vector with equation \eqref{21}, so that the invariant information can guide the update of vector information.

\subsection{Optimization Objective}
In this section, we will introduce two loss functions for protein-ligand distance matrix, and self-confidence, which are inspired by the previous method \cite{lutankbind}.

\subsubsection{Protein-Ligand distance matrix}
We utilize the previous methods to predict the distance matrix $\hat{D}_{ij}$ between proteins and ligands, which can then be used to generate the docking pose \cite{lutankbind, jumper2021highly}. After processing the features of the ligand and protein using feature processing  and TAMformer, we use MLP to predict the final distance matrix $\hat{D}_{ij}$:
\begin{equation}
\label{predict_dis}
\hat{D}_{ij} = MLP(\z_{ij}),
\end{equation}
and then compute a loss as
\begin{equation}
\label{}
\mathcal{L}_{a}=\delta(r)\frac{1}{sq}\sum_{i=1}^{s}\sum_{j=1}^{q} \left \| \hat{D}_{ij} - D_{ij} \right \|,
\end{equation}
where $D_{ij}$ is the true distance between ligand and protein. $r$ is the pocket number by P2Rank. In addition, we utilize root mean square error (RMSE) to minimize the difference between the predicted distance matrix $\hat{D}_{ij}$ and the true distance matrix $D_{ij}$ of the protein and ligand.
$\delta(r)$ is set to 1 when it is close to the native pocket, and 0 otherwise. 

\subsubsection{Self-confidence} 
In protein-ligand docking, there may be multiple potential binding pockets on the protein surface. To address this challenge, we use the P2Rank algorithm to generate a list of the top ten possible binding pockets on the protein surface\cite{krivak2018p2rank}. However, after the ligand binds to one of the pockets, the likelihood of the ligand binding to other pockets is greatly reduced. To account for this, we employ a self-confidence function $\mathcal{L}_{b}$, which considers the probability of the ligand binding to each of the ten predicted pockets:
\begin{equation}
\label{}
\hat{f}_{r} = \sum_{i=1}^{s}\sum_{j=1}^{q}MLP(\z_{i j}),
\end{equation}
\begin{equation}
\label{}
\mathcal{L}_{b}(\hat{f}_{r},f)= \\
\delta(r)(\hat{f}_{r}-f)^{2} + (1-\delta (r))max(0,\hat{f}_{r}-(f-\epsilon))^2,
\end{equation}
where $\hat{f}_{r}$ is the confident score of pocket $r$ and $\epsilon$ is the margin value. We utilize binding affinity as the true label $f$. Our final optimization objective $\mathcal{L}$ is  the sum of $\mathcal{L}_{a}$ and  $\mathcal{L}_{b}$:
\begin{equation}
\label{}
\mathcal{L}= \mathcal{L}_{a} + \alpha\mathcal{L}_{b},
\end{equation}
where $\alpha$ is a hyperparameter, which is set to 1 in our experiments.
\begin{table*}[]
\center
\large
\caption{Experimental results on the PDBbind v2020 dataset}
\label{result}
\begin{tabular}{@{}lllllllllllll@{}}
\toprule
         & \multicolumn{6}{c}{\textbf{LIGAND RMSD}}                                                                                   & \multicolumn{6}{c}{\textbf{CENTROID DISTANCE}}                                                                             \\
         & \multicolumn{4}{c}{Percentiles $\downarrow$} & \multicolumn{2}{c}{\begin{tabular}[c]{@{}c@{}}\%Below\\ Threshold $\uparrow$\end{tabular}} & \multicolumn{4}{c}{Percentiles $\downarrow $} & \multicolumn{2}{l}{\begin{tabular}[c]{@{}l@{}}\%Below\\ Threshold $\uparrow$\end{tabular}} \\ \cmidrule(lr){2-5} \cmidrule(lr){8-11}
\textbf{Methods}  & 25\%   & 50\%   & 75\%  & Mean  & 2\AA                                     & 5\AA                                     & 25\%   & 50\%   & 75\%  & Mean  & 2\AA                                     & 5\AA                                     \\ \midrule
VINA     & 5.7    & 10.7   & 21.4  & 14.7  & 5.5                                    & 21.2                                   & 1.9    & 6.2    & 20.1  & 12.1  & 26.5                                   & 47.1                                   \\ 
SMINA    & 3.8    & 8.1    & 17.9  & 12.1  & 13.5                                   & 33.9                                   & 1.3    & 3.7    & 16.2  & 9.8   & 38.0                                   & 55.9                                   \\
QVINA-W  & 2.5    & 7.7    & 23.7  & 13.6  & 20.9                                   & 40.2                                   & 0.9    & 3.7    & 22.9  & 11.9  & 41.0                                   & 54.6                                   \\
GNINA    & 2.8    & 8.7    & 22.1  & 13.3  & 21.2                                   & 37.1                                   & 1.0    & 4.5    & 21.2  & 11.5  & 36.0                                   & 52.0                                   \\
GLIDE    & 2.6    & 9.3    & 28.1  & 16.2  & 21.8                                   & 33.6                                   & 0.8    & 5.6    & 26.9  & 14.4  & 36.1                                   & 48.7                                   \\ \midrule
EquiBind & 3.9    & 6.3    & 10.5  & 8.3   & 4.1                                    & 40.2                                   & 1.2    & 2.7    & 6.9   & 5.5   & 40.2                                  & 69.7                                   \\
TankBind & 2.5    & 4.4    & 8.4   & 7.9   & 19.0                                   & 56.4                                   & 0.8    & 1.7    & 4.4   & 5.8   & 55.3                                   & 77.4                                   \\ \midrule
\textbf{ETDock}  & \textbf{2.1}    & \textbf{3.8}    & \textbf{7.7}   & \textbf{7.4}   & \textbf{23.2}                                   & \textbf{61.1}                                   & \textbf{0.7}    & \textbf{1.4}    & \textbf{3.8}   & \textbf{5.5}   & \textbf{59.1}                                   & \textbf{79.3}                                   \\ \bottomrule
\end{tabular}
\end{table*}

\subsection{Generate the Binding Ligand Pose}
\begin{figure}[h]
\centering
\includegraphics[width=1.0\linewidth]{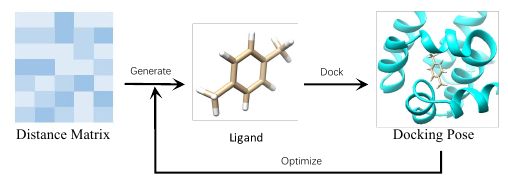} 
\caption{The workflow of generate the binding ligand pose. }
\label{fig:generate}
\end{figure}
Our model has the capability to predict the distance map between the protein and ligand, so we employ a two-stage approach to reconstruct the three-dimensional structure of the ligand based on the protein-ligand  distance map. Inspired by previous methods\cite{lutankbind, jumper2021highly}, we iteratively generate the final ligand binding structure by incorporating distance constraints within the ligand  $\mathcal{L}_{inter}$ and leveraging local atomic structure constraints $\mathcal{L}_{stru}$:
\begin{equation}
\label{}
\mathcal{L}_{inter} = \sum_{i=1}^{s}\sum_{j=1}^{q}(|\Check{D}_{ij} - \hat{D}_{ij}|),
\end{equation}
\begin{equation}
\label{}
\mathcal{L}_{stru} = \sum_{j=1}^{s}\sum_{k=1}^{s}(|\Check{D}_{jk}^{l} - D_{jk}^{l}|),
\end{equation}
where $\Check{D}_{ij}=||\hat{c}_{i}^{l}-c_{j}^{p}||$  and $\Check{D}_{jk}^{l}=||\hat{c}_{j}^{l}-\hat{c}_{k}^{l}||$. $c_{i}^{p}$ are the coordinates of protein nodes. $\hat{c}_{i}^{l}$ represents the final predicted coordinates of the ligand atoms. $D_{jk}^{l}$ represents the distance matrix between pairs of atoms coordinates within the ligand, which are obtained by RDKit. The generating objective $\mathcal{L}_{generate}$ can be formulated as follows:
\begin{equation}
\label{}
\mathcal{L}_{generate} = (1-\beta)\mathcal{L}_{inter} + \beta \mathcal{L}_{stru}.
\end{equation}
where $\beta$ is the hyperparameter. The workflow to generate the binding ligand pose is demonstrated in Fig. \ref{fig:generate}.

\section{Experiments}

\subsection{Experimental Settings}
\subsubsection{Dataset} We assess the performance of our model in protein-ligand docking on the PDBbind dataset \cite{liu2015pdb}. The PDBbind dataset includes structural data collected from the Protein Data Bank (PDB) along with associated experimental measurements \cite{burley2021rcsb}. The PDBbind dataset also provides structural information about protein-ligand complexes, including atomic coordinates of proteins and structural and chemical information of ligands. Additionally, the PDBbind dataset contains experimentally determined binding affinities for protein-ligand complexes. We use PDBbind v2020 which has 19443 protein-ligand complexes and adopt the time split strategy described in the EquiBind \cite{stark2022equibind}, i.e., our dataset is split based on the deposition year of protein-ligand complex structures. The training and validation sets include structures deposited before 2019, while the test set consists of structures deposited after 2019. By eliminating a subset of structures that could not be processed using RDKit, our training set consists of 17,787 structures \cite{landrum2013rdkit}. For the purpose of validation, we allocate 968 structures, while 363 structures are designated for testing.

\subsubsection{Evaluation Metrics}
Following previous studies \cite{lutankbind, stark2022equibind}, we evaluate the performance of our model using root mean square deviation (RMSD). In particular, we calculate RMSD between the predicted and true positions of ligand atoms, and assess the accuracy of ligand pose prediction. Additionally, we employ RMSD between the centroid distances of the ligand and the true distances to measure the ability of our model in identifying the correct binding region. Furthermore, we utilize quantiles and mean values to assess the predictive performance of our model across different result ranges. Considering that an RMSD below 2\AA \space is deemed acceptable, we calculate the percentage of predicted ligand poses with an RMSD below 2\AA \space and below 5\AA. This analysis provides additional evidence of the ability of our model in predicting ligand poses. Note that when calculating these metrics, we exclude hydrogen atoms from the ligand.

\subsubsection{Baselines} 
We compare our ETDock with state-of-the-art baselines from 2 categories as follows:

\textbf{Search-based docking methods:}
\begin{itemize}
    \item  AutoDock \textbf{VINA} is a popular molecular docking software that utilizes an efficient search algorithm and scoring function to explore the conformational space of ligands and predict their binding modes to target proteins\cite{trott2010autodock}. 

    \item \textbf{SMINA} is an enhanced molecular docking software derived from AutoDock VINA, incorporating optimized search algorithms and scoring functions\cite{koes2013lessons}. 
    
    \item \textbf{GNINA} enhances SMINA by incorporating a learned 3D Convolutional Neural Network for scoring\cite{mcnutt2021gnina}. 
    
    \item \textbf{QVINA-W} is a blind docking software that builds upon the speed-optimized QuickVina 2 by incorporating advanced algorithms for efficient exploration of ligand binding modes\cite{hassan2017protein}. 
    
    \item \textbf{GLIDE} docking approach includes initial rough positioning, torsionally flexible energy optimization, and Monte Carlo sampling to obtain precise ligand
     docking.\cite{friesner2004glide}.
\end{itemize}

\textbf{Deep learning docking methods:}
\begin{itemize}
    \item \textbf{EquiBind} is an equivariant model that directly predicts the binding pose structure coordinates. It refines the ligand conformation using graph neural networks and aligns the refined ligand by keypoint alignment mechanism into the binding pocket\cite{stark2022equibind}. 
    
    \item \textbf{Tankbind} employs a deep learning approach to predict the protein-ligand distance matrix, enabling the generation of docking pose. By leveraging optimization algorithms, it reduces the reliance on exhaustive conformational sampling, improving computational efficiency in ligand binding prediction\cite{lutankbind}.
\end{itemize}

\subsubsection{Implementation Details}
We trained our model using the Adam optimizer with a learning rate of 0.0001 for a total of 500 epochs. The model with the highest validation score was selected and subsequently evaluated on the independent test set. The training process was performed on single Tesla V100 GPU 32G and
Inter Xeon Gold 5218 16-Core Processor.  The number of layers in TAMformer is customizable. In our model, we utilized 3 layers. The dataset partitioning was based on the strategy employed in EquiBind\cite{stark2022equibind}. To prevent label leakage, we utilized RDKit to generate input coordinates for the ligands. We used TorchDrug to extract atomic features for the ligands\cite{zhu2022torchdrug}. The dimensions of the node embeddings for both proteins and ligands were set to 128. For identifying functional regions, we employed P2Rank to identify the top 10 potential binding site regions in the protein.   Additionally, in ligand pose generation based on the distance matrix, we employed  the Adam optimizer with a learning rate of 0.2, conducting 8000 iterations.
\begin{table}[h]
\caption{Ablation Studies.}
\label{tab2}
\begin{tabular}{@{}lccccc@{}}
\toprule
                  & \textbf{\begin{tabular}[c]{@{}c@{}}Feature\\ fusion\end{tabular}} & \textbf{\begin{tabular}[c]{@{}c@{}}Triangle\\ Layer\end{tabular}} & \textbf{\begin{tabular}[c]{@{}c@{}}Attention\\ Layer\end{tabular}} & \textbf{\begin{tabular}[c]{@{}c@{}}Message\\ Layer\end{tabular}} & \textbf{\begin{tabular}[c]{@{}c@{}}Equivariant\\ vector\end{tabular}} \\ \midrule
\textbf{ETDock-F} & $ {} $ {\color{red}\XSolidBrush}                                                                & \Checkmark                                                                   & \Checkmark                                                                    & \Checkmark                                                                   & \Checkmark                                                                        \\ \midrule
\textbf{ETDock-T} & \Checkmark                                                               & {\color{red}\XSolidBrush}                                                                 & \Checkmark                                                                     & \Checkmark                                                                   & \Checkmark                                                                        \\ \midrule
\textbf{ETDock-A} & \Checkmark                                                                    & \Checkmark                                                                    & {\color{red}\XSolidBrush}                                                                      & \Checkmark                                                                   & \Checkmark                                                                        \\ \midrule
\textbf{ETDock-M} &  \Checkmark                                                                   & \Checkmark                                                                    & \Checkmark                                                                    & {\color{red}\XSolidBrush}                                                                  & {\color{red}\XSolidBrush}                                                                        \\ \midrule
\textbf{ETDock-E} & \Checkmark                                                                    & \Checkmark                                                                    &  \Checkmark                                                                   &  \Checkmark                                                                  & {\color{red}\XSolidBrush}                                                                       \\ \bottomrule
\end{tabular}
\end{table}
\begin{table*}[]
\center
\large
\caption{The experimental results of ablation studies.}
\label{tab3}
\begin{tabular}{@{}lllllllllllll@{}}
\toprule
         & \multicolumn{6}{c}{\textbf{LIGAND RMSD}}                                                                                   & \multicolumn{6}{c}{\textbf{CENTROID DISTANCE}}                                                                             \\
         & \multicolumn{4}{c}{Percentiles $\downarrow$} & \multicolumn{2}{c}{\begin{tabular}[c]{@{}c@{}}\%Below\\ Threshold $\uparrow$\end{tabular}} & \multicolumn{4}{c}{Percentiles $\downarrow $} & \multicolumn{2}{l}{\begin{tabular}[c]{@{}l@{}}\%Below\\ Threshold $\uparrow$\end{tabular}} \\ \cmidrule(lr){2-5} \cmidrule(lr){8-11}
\textbf{Methods}  & 25\%   & 50\%   & 75\%  & Mean  & 2\AA                                     & 5\AA                                     & 25\%   & 50\%   & 75\%  & Mean  & 2\AA                                     & 5\AA                                     \\ \midrule
ETDock-F  & 2.3    & 4.1    & 8.1  & 7.7  & 19.6                                   & 58.9                                   & 0.8    & 1.7    & 4.2  & 5.7  & 55.9                                 & 77.1                                   \\
ETDock-T    & 2.7   & 4.8   & 8.3  & 8.4  & 15.1                                   & 52.8                                   & 0.8    & 1.7    & 5.2  & 6.1  & 53.9                                 & 74.1                                  \\
ETDock-A    & 2.4    & 4.2   & 7.5  & 8.0  & 18.4                                  & 58.1                                   & 0.7    & 1.7    & 4.1  & 6.0  & 55.6                                   & 78.7                                   \\
ETDock-M     & 3.3   & 5.3   & 9.1  & 8.7  & 9.0                                    & 46.8                                   & 1.2    & 2.7    & 6.3  & 6.7  & 40.2                                  & 68.0                                   \\
ETDock-E    & 2.6    & 4.4    & 8.4  & 8.8  & 18.1                                   & 55.9                                   & 0.8    & 1.6    & 5.0  & 6.8   & 56.1                                 & 75.0                                 \\
ETDock   & \textbf{2.1}    & \textbf{3.8}    & \textbf{7.7}   & \textbf{7.4}   & \textbf{23.2}                                   & \textbf{61.1}                                   & \textbf{0.7}    & \textbf{1.4}    & \textbf{3.8}   & \textbf{5.5}   & \textbf{59.1}                                   & \textbf{79.3}                                   \\ \bottomrule
\end{tabular}
\end{table*}

\subsection{Performance Evaluation}
\subsubsection{Overall Comparison} 
Table \ref{result} shows the performance of our ETDock method compared to other baselines on the PDBbind v2020 test dataset. ETDock outperforms existing search-based docking methods and deep learning docking methods. The percentage of results with ligand RMSD below 2 \AA \space is 23.3\%, while the percentage of results below 5 \AA \space is 61.1\%.

VINA utilizes a scoring function to perform a search in the ligand conformational space for ligand-protein docking. SMINA improves Vina's optimization algorithm, resulting in an increase in the percentage of ligand RMSD below 2 \AA \space from 5.5\% to 13.5\%. GNINA employs a 3D CNN as a scoring function, allowing it to capture more comprehensive spatial information for enhanced ligand-protein docking. GNINA achieves a significant performance improvement, with the percentage of ligand RMSD below 2 \AA \space increasing from 13.5\% to 21.2\%. QVINA-W significantly enhances the speed of the optimization algorithm. GLIDE incorporates energy-based optimization, enabling it to achieve optimal results among search-based methods. Equibind, the first method to employ deep learning for ligand-protein docking, drastically reduces the docking time but fails to surpass the accuracy of optimal search-based methods\cite{stark2022equibind}. Tankbind incorporates physical constraints and contrastive learning methods, leading to improved performance. The percentage of ligand RMSD below 2 \AA \space increases from 4.1\% to 19\%, representing substantial progress\cite{lutankbind}. In comparison, our ETDock method leverages Transformer-based models to consider equivariant information, outperforming all previous methods in terms of performance on the ligand-protein docking task.

\subsubsection{Ablation Studies} 
In ETDock, we have designed five modules, including feature fusion, the triangle layer, the attention layer, the message layer and equivariant vector. To further demonstrate the effectiveness of these modules within our model, we have developed the following five ETDock variants in Table \ref{tab2}:
a. \textbf{ETDock-F}: In this variant, we remove the feature fusion module from the ETDock architecture.
b. \textbf{ETDock-T}: we remove the triangle layer module from the ETDock  to validate the effectiveness of the physical constraints.
c. \textbf{ETDock-A}: we remove the attention layer module from the ETDock to assess the ability of the attention layer to capture separate information from the ligand-protein interaction features.
d. \textbf{ETDock-M}: we remove the message layer module from the ETDock to validate the necessity of the message layer for the interaction between invariant and equivariant features.
e. \textbf{ETDock-E}: we remove the equivariant vector feature learning component from the message layer in the ETDock to assess the impact of vector features on message passing, while keeping the other components within the message layer unchanged.

In the aforementioned variants, each modification only affects one module while keeping the other modules in ETDock unchanged. The experimental results of the five variants are shown in Table \ref{tab3}.

According to the results shown in Table \ref{tab3}, we observe that the ETDock-M performs the worst, indicating that the interaction between invariant and equivariant vector information in the message layer is crucial for our task. Additionally, the ETDock-E suggests that the equivariant vector information in the message layer contributes significantly to the overall  performance. These experimental results also validate the significance of 3D spatial information in predicting protein-ligand docking. The results of the ETDock-T demonstrate the importance of capturing the physical constraints between ligands and proteins for effective feature learning. The performance of ETDock-F highlights the effectiveness of integrating atom-level and graph-level features of the ligand, enabling a more comprehensive learning of the ligand's information. 
ETDock-A shows that the attention layer can effectively promote the learning of ligand features and protein features from interactive features.

\section{analysis}
\subsection{Hyperparameter}
In generating ligand poses through distance matrix, the parameter $\beta$ in equation 29 has a significant impact. We conducted a parameter analysis on $\beta$, as shown in Figure 3.
\begin{figure}[h]
\centering
\includegraphics[width=0.95\linewidth]{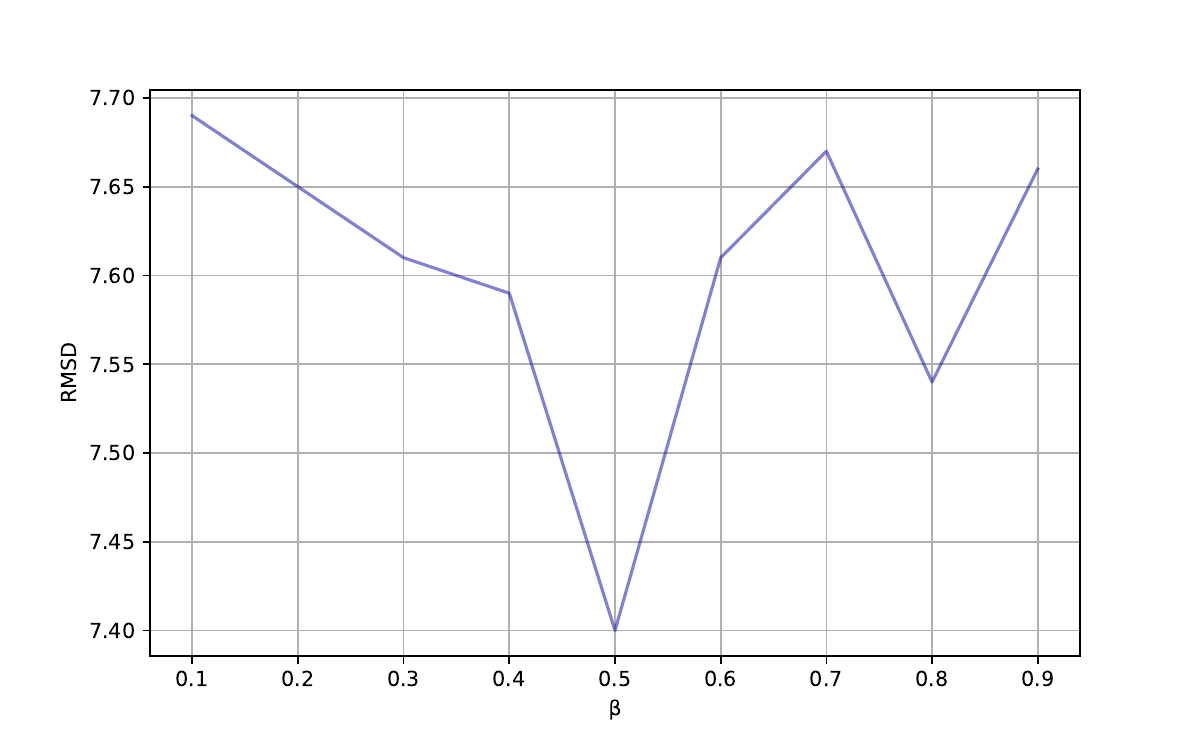} 
\caption{The experimental results on the hyperparameter $\beta$.}
\end{figure}

$\beta$ determines the weighting between two loss functions, $\mathcal{L}_{inter}$ and $\mathcal{L}_{stru}$. From Figure 3, it can be observed that as the weight of $\mathcal{L}_{stru}$ increases, the RMSD of the ligand pose decreases continuously. When $\beta$ is set to 0.5, the ligand pose achieves the minimum RMSD.
When $\beta$ is increased beyond 0.5, with the weight of $\mathcal{L}_{stru}$ surpassing that of $\mathcal{L}_{inter}$, it has been observed that the performance of ligand pose generation tends to decline. This suggests that the incorporation of distance constraints within the ligand becomes increasingly valuable in the generation process. By assigning a higher weight to $\mathcal{L}_{inter}$, which represents the preservation of distance relationships, the algorithm can better capture and maintain the spatial arrangement of the ligand atoms. This emphasis on distance constraints aids in generating ligand poses with improved accuracy and alignment to the target structure.

\begin{figure}[h]
\centering
\includegraphics[width=0.95\linewidth]{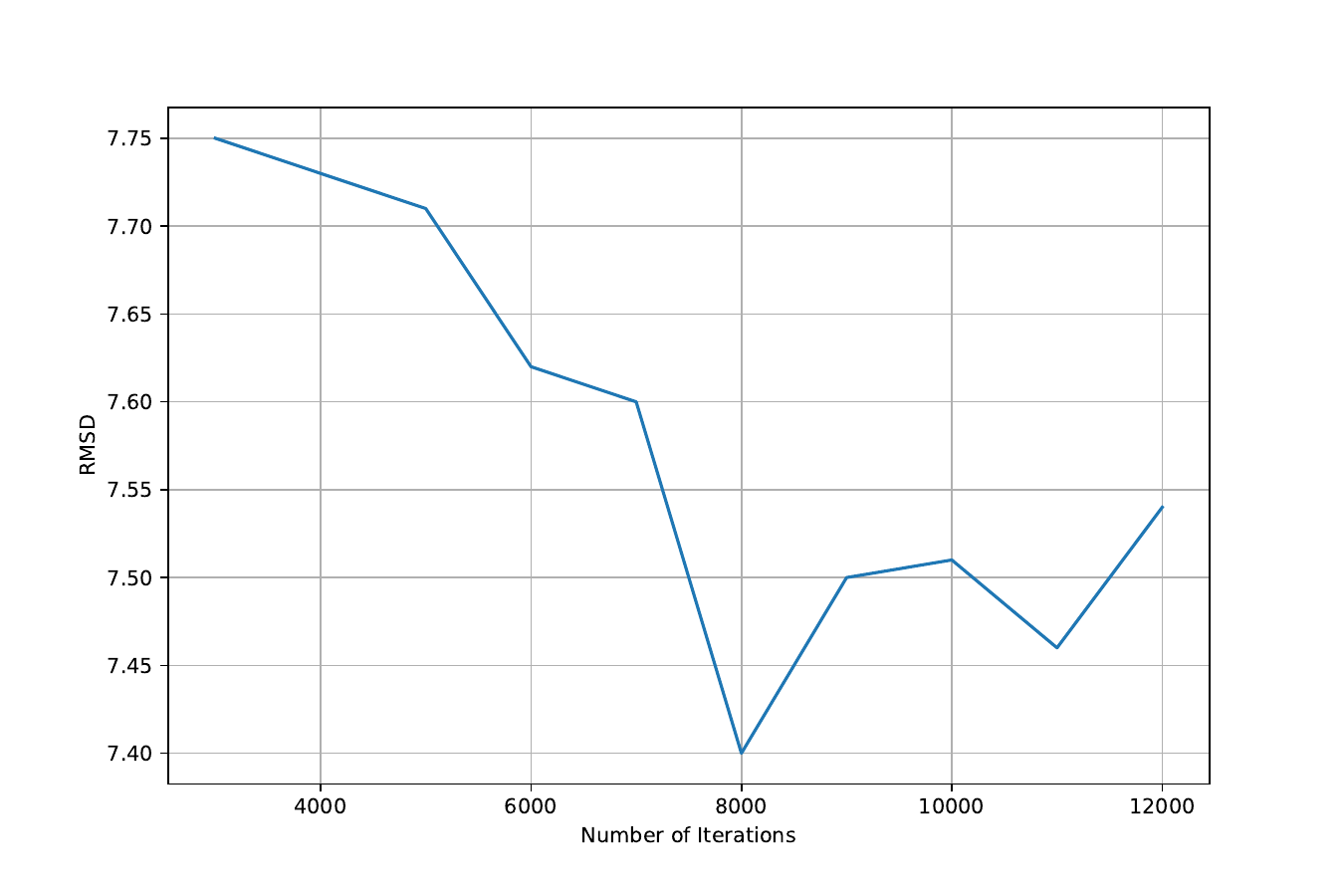} 
\caption{The experimental results on number of iterations during ligand pose generation. We utilize the average RMSD of the ligand to assess the impact of the number of iterations in the optimization algorithm.}
\label{fig4}
\end{figure}

During the ligand pose generation phase, we employ an iterative optimization algorithm based on the distance matrix. The number of iterations plays a crucial role in determining the precision of the resulting ligand pose and directly influences the computational time required for the process. Consequently, we conducted experimental analyses varying the number of iterations to assess its impact on the quality of the generated ligand poses in Figure \ref{fig4}.

From Figure \ref{fig4}, we observe a consistent decrease in the average RMSD of the ligand as the number of iterations increases. This indicates that with an increasing number of iterations, the ligand pose can be optimized to a more precise position. The minimum average RMSD for the ligand is achieved when the number of iterations reaches 8000. However, when we further increase the number of iterations, the average rmsd of the ligand starts to increase again. This suggests that the optimization process may become overly fine-tuned, resulting in diminishing returns.Considering both the diminishing improvement in rmsd and the increasing computational time as the number of iterations grows, we select 8000 iterations as our final choice. This balance allows us to achieve a sufficiently accurate ligand pose optimization without excessive computational overhead.

\begin{figure*}[]
\centering
\includegraphics[width=0.95\linewidth]{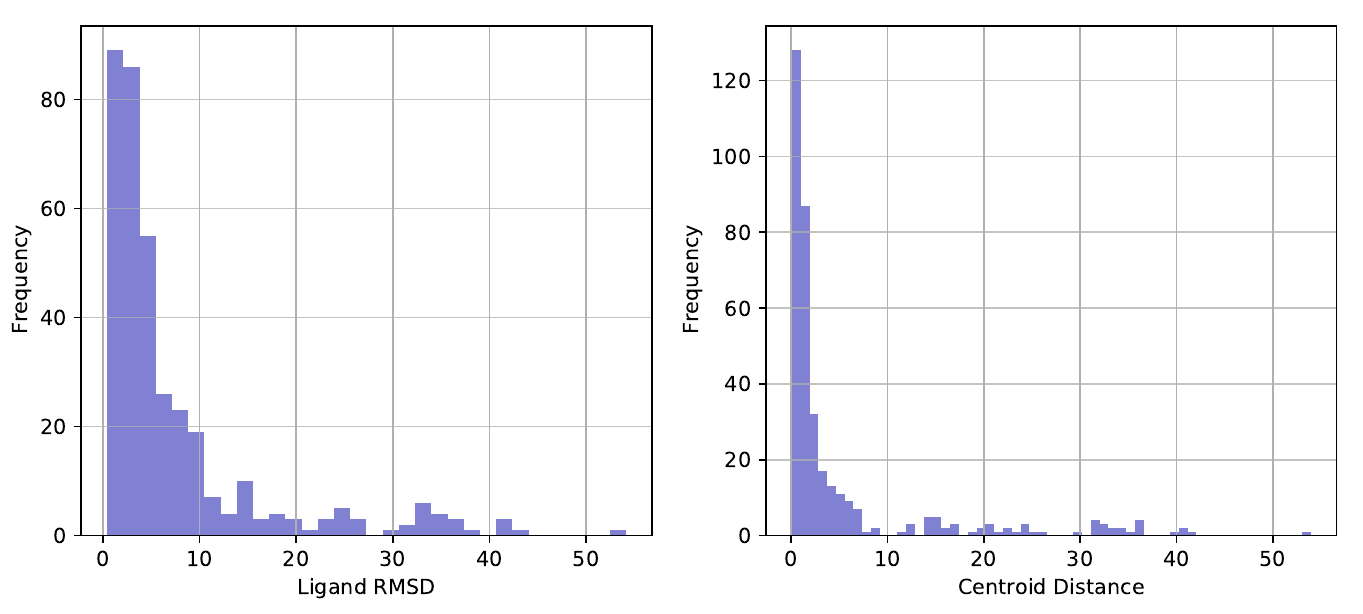} 
\caption{The frequency histograms of Ligand RMSD(left) and Centeroid Distances(right) predicted by ETDock on the test set. These histograms provide a visual representation of the distribution and occurrence frequency of the RMSD and the centeroid distance for the predicted ligands,  }
\label{frequency}
\end{figure*}

\subsection{Visualization}
\begin{figure*}[]
\centering
\includegraphics[width=0.9\linewidth]{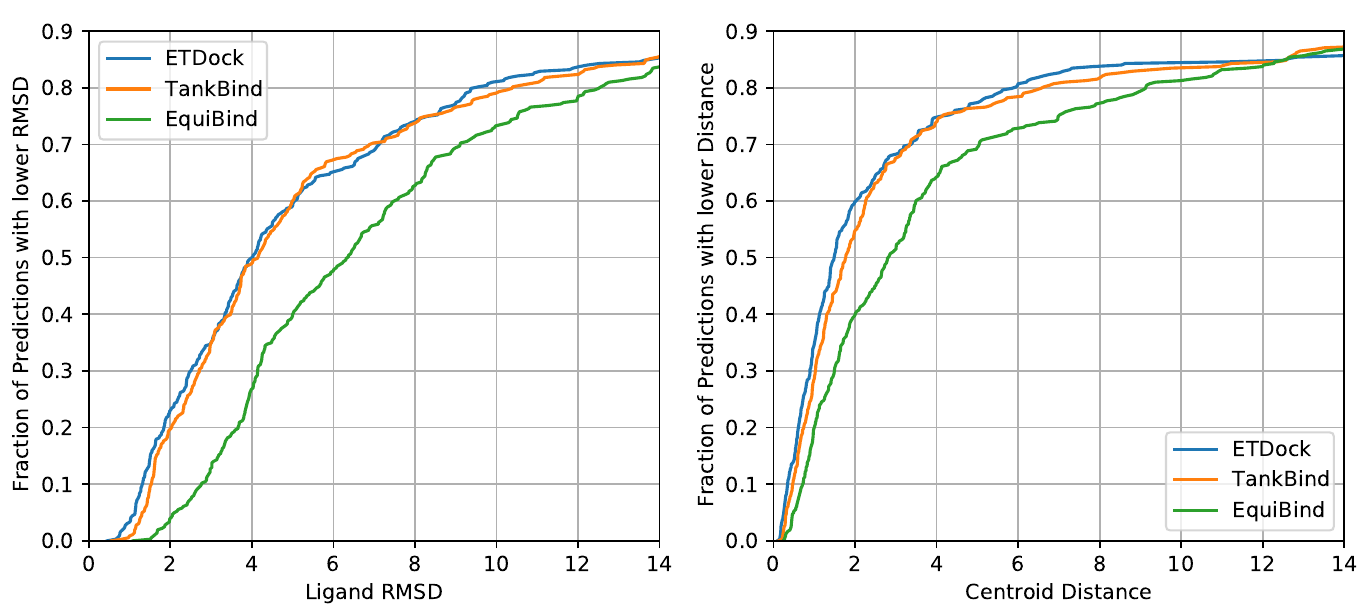} 
\caption{Estimator of the Cumulative Distribution Function (ECDF) plot for ligand RMSD (left) and Centroid Distance (right) from result evaluated on the test dataset. }
\label{ecdf}
\end{figure*}
In Table \ref{result}, the experimental results are presented using mean and quantiles, which may not provide a clear understanding of the frequency distribution of predicted the ligand RMSD and the centroid distance. To address this issue, we computed the RMSD and the centroid distance between the predicted ligand coordinates generated by ETDock on the test set and the corresponding true labels. Subsequently, we utilized frequency histograms to visualize the distribution of these RMSD and the centroid distance in Fig. \ref{frequency}.


In Fig. \ref{frequency}, the ligand RMSD plot showcases the performance of ETDock in predicting ligand conformational accuracy. It can be observed that out of the 363 ligands evaluated, more than 80 ligands have a predicted RMSD from the true label that falls below the threshold of 2 Å. This indicates that ETDock demonstrates a high capability in accurately predicting ligand conformations, with a significant number of ligands achieving a level of precision within the 2 Å range.
Furthermore, the centroid distance plot reveals that over 120 ligands exhibit a predicted centroid distance of less than 2 Å. The centroid distance metric assesses the accuracy of the predicted ligand centroid position compared to the true centroid position. The fact that a considerable number of ligands achieve a centroid distance below 2 Å highlights the ability of ETDock to accurately predict the overall position of the ligand within the binding site.
These results signify the effectiveness of ETDock in generating ligand poses with a high level of accuracy, as evidenced by the low RMSD values and centroid distances.

In ligand-protein docking tasks, we are particularly interested in the percentage of predicted ligand coordinates that have the RMSD and the centroid distance less than 2 \AA \space compared to the true label. To gain a better understanding of these percentages, we visualize them using the estimator of the cumulative distribution function (ECDF), which provide a clear representation of the distribution of values in Fig. \ref{ecdf}.


In Fig. \ref{ecdf}, the ECDF curves of ETDock for ligand RMSD and centroid distance can be observed. Notably, these ECDF curves envelop the curves of Tankbind and Equibind, particularly when considering values below 2 Å. This suggests that ETDock surpasses Tankbind and Equibind in terms of predictive performance, specifically for values below 2 Å.
The fact that the ECDF curves of ETDock enclose those of Tankbind and Equibind implies that ETDock exhibits a higher proportion of ligand poses with lower RMSD and centroid distance values compared to the other methods. This indicates that ETDock has a superior ability to generate ligand poses that closely resemble the true conformation, especially for values below the 2 Å threshold. These findings demonstrate the improved predictive performance of ETDock in accurately predicting ligand conformations within this specific range.

\subsection{Cases Study}
\begin{figure}[h]
\centering
\includegraphics[width=0.9\linewidth]{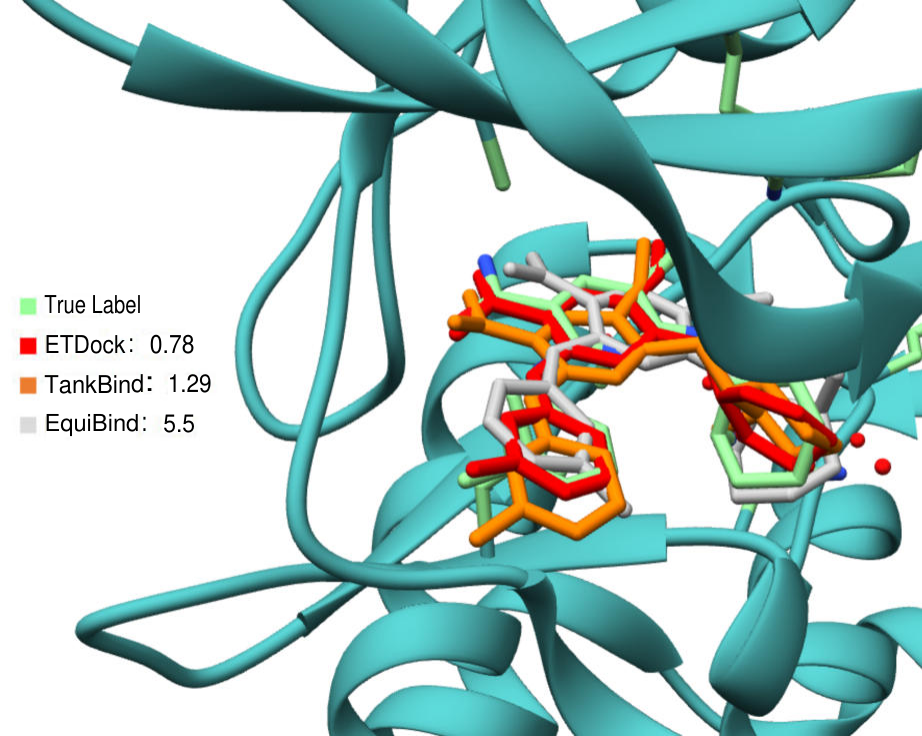} 
\caption{Case study on PDB 6JMF. The figure presents the predicted poses generated by ETDock (red), TankBind (orange), and EquiBind (gray) in relation to the target protein. The RMSD values, indicating the dissimilarity between the predicted poses and the true label ligand poses (green), are also presented.}
\label{case}
\end{figure}

Moreover, we performed a comparative analysis of ligand pose predictions between ETDock, TankBind, and EquiBind on the test dataset (6JMF) in Fig. \ref{case}. Through a case study on carefully selected representative samples, we observed that ETDock exhibited a remarkably lower RMSD of 0.78 \AA, compared to the other two state-of-the-art methods. This result highlights the superior predictive accuracy of ETDock in capturing the true conformation of the ligand.

\section{Conclusion}
In this paper, we propose a novel and high-performance equivariant transformer for predicting protein-ligand docking. Our model leverages invariant  information and equivariant vector information to learn embeddings of ligands and proteins. Specifically, we introduce a feature processing module to encode the features of ligands and proteins. Subsequently, we introduce a TAMformer module to learn both scalar and equivariant vector information of ligands and proteins. Experimental results on the PDBbind v2020 dataset demonstrate that our ETDock outperforms previous traditional and deep learning methods, achieving state-of-the-art performance.

\section*{REFERENCES}
\bibliographystyle{IEEEtran}
\bibliography{mybib}

\begin{thebibliography}{10}
\providecommand{\url}[1]{#1}
\csname url@samestyle\endcsname
\providecommand{\newblock}{\relax}
\providecommand{\bibinfo}[2]{#2}
\providecommand{\BIBentrySTDinterwordspacing}{\spaceskip=0pt\relax}
\providecommand{\BIBentryALTinterwordstretchfactor}{4}
\providecommand{\BIBentryALTinterwordspacing}{\spaceskip=\fontdimen2\font plus
\BIBentryALTinterwordstretchfactor\fontdimen3\font minus
  \fontdimen4\font\relax}
\providecommand{\BIBforeignlanguage}[2]{{%
\expandafter\ifx\csname l@#1\endcsname\relax
\typeout{** WARNING: IEEEtran.bst: No hyphenation pattern has been}%
\typeout{** loaded for the language `#1'. Using the pattern for}%
\typeout{** the default language instead.}%
\else
\language=\csname l@#1\endcsname
\fi
#2}}
\providecommand{\BIBdecl}{\relax}
\BIBdecl

\bibitem{du2021trrosetta}
Z.~Du, H.~Su, W.~Wang, L.~Ye, H.~Wei, Z.~Peng, I.~Anishchenko, D.~Baker, and
  J.~Yang, ``The trrosetta server for fast and accurate protein structure
  prediction,'' \emph{Nature protocols}, vol.~16, no.~12, pp. 5634--5651, 2021.

\bibitem{jumper2021highly}
J.~Jumper, R.~Evans, A.~Pritzel, T.~Green, M.~Figurnov, O.~Ronneberger,
  K.~Tunyasuvunakool, R.~Bates, A.~{\v{Z}}{\'\i}dek, A.~Potapenko
  \emph{et~al.}, ``Highly accurate protein structure prediction with
  alphafold,'' \emph{Nature}, vol. 596, no. 7873, pp. 583--589, 2021.

\bibitem{jin2022antibody}
W.~Jin, R.~Barzilay, and T.~Jaakkola, ``Antibody-antigen docking and design via
  hierarchical structure refinement,'' in \emph{International Conference on
  Machine Learning}.\hskip 1em plus 0.5em minus 0.4em\relax PMLR, 2022, pp.
  10\,217--10\,227.

\bibitem{giri2020multipredgo}
S.~J. Giri, P.~Dutta, P.~Halani, and S.~Saha, ``Multipredgo: deep multi-modal
  protein function prediction by amalgamating protein structure, sequence, and
  interaction information,'' \emph{IEEE Journal of Biomedical and Health
  Informatics}, vol.~25, no.~5, pp. 1832--1838, 2020.

\bibitem{zhou2022tasser}
X.~Zhou, W.~Zheng, Y.~Li, R.~Pearce, C.~Zhang, E.~W. Bell, G.~Zhang, and
  Y.~Zhang, ``I-tasser-mtd: a deep-learning-based platform for multi-domain
  protein structure and function prediction,'' \emph{Nature Protocols},
  vol.~17, no.~10, pp. 2326--2353, 2022.

\bibitem{fowler2022accuracy}
N.~J. Fowler and M.~P. Williamson, ``The accuracy of protein structures in
  solution determined by alphafold and nmr,'' \emph{Structure}, vol.~30, no.~7,
  pp. 925--933, 2022.

\bibitem{sadybekov2023computational}
A.~V. Sadybekov and V.~Katritch, ``Computational approaches streamlining drug
  discovery,'' \emph{Nature}, vol. 616, no. 7958, pp. 673--685, 2023.

\bibitem{nussinov2023alphafold}
R.~Nussinov, M.~Zhang, Y.~Liu, and H.~Jang, ``Alphafold, allosteric, and
  orthosteric drug discovery: Ways forward,'' \emph{Drug Discovery Today}, p.
  103551, 2023.

\bibitem{liu2023efficient}
H.~Liu, C.~Wang, P.~Liu, C.~Liu, Z.~Wang, and Z.~Wei, ``Efficient large-scale
  virtual screening based on heterogeneous many-core supercomputing system,''
  \emph{IEEE Journal of Biomedical and Health Informatics}, 2023.

\bibitem{10027686}
Y.~Zhang, G.~Zhou, Z.~Wei, and H.~Xu, ``Predicting protein-ligand binding
  affinity via joint global-local interaction modeling,'' in \emph{2022 IEEE
  International Conference on Data Mining (ICDM)}, 2022, pp. 1323--1328.

\bibitem{ma2023predicting}
W.~Ma, S.~Zhang, Z.~Li, M.~Jiang, S.~Wang, N.~Guo, Y.~Li, X.~Bi, H.~Jiang, and
  Z.~Wei, ``Predicting drug-target affinity by learning protein knowledge from
  biological networks,'' \emph{IEEE Journal of Biomedical and Health
  Informatics}, vol.~27, no.~4, pp. 2128--2137, 2023.

\bibitem{hua2023mfr}
Y.~Hua, X.~Song, Z.~Feng, and X.~Wu, ``Mfr-dta: a multi-functional and robust
  model for predicting drug--target binding affinity and region,''
  \emph{Bioinformatics}, vol.~39, no.~2, p. btad056, 2023.

\bibitem{li2021structure}
S.~Li, J.~Zhou, T.~Xu, L.~Huang, F.~Wang, H.~Xiong, W.~Huang, D.~Dou, and
  H.~Xiong, ``Structure-aware interactive graph neural networks for the
  prediction of protein-ligand binding affinity,'' in \emph{Proceedings of the
  27th ACM SIGKDD Conference on Knowledge Discovery \& Data Mining}, 2021, pp.
  975--985.

\bibitem{lu2023improving}
R.~Lu, J.~Wang, P.~Li, Y.~Li, S.~Tan, Y.~Pan, H.~Liu, P.~Gao, G.~Xie, and
  X.~Yao, ``Improving drug-target affinity prediction via feature fusion and
  knowledge distillation,'' \emph{Briefings in Bioinformatics}, vol.~24, no.~3,
  p. bbad145, 2023.

\bibitem{chu2022hierarchical}
Z.~Chu, F.~Huang, H.~Fu, Y.~Quan, X.~Zhou, S.~Liu, and W.~Zhang, ``Hierarchical
  graph representation learning for the prediction of drug-target binding
  affinity,'' \emph{Information Sciences}, vol. 613, pp. 507--523, 2022.

\bibitem{yang2023geometric}
Z.~Yang, W.~Zhong, Q.~Lv, T.~Dong, and C.~Yu-Chian~Chen, ``Geometric
  interaction graph neural network for predicting protein--ligand binding
  affinities from 3d structures (gign),'' \emph{The Journal of Physical
  Chemistry Letters}, vol.~14, no.~8, pp. 2020--2033, 2023.

\bibitem{lutankbind}
W.~Lu, Q.~Wu, J.~Zhang, J.~Rao, C.~Li, and S.~Zheng, ``Tankbind:
  Trigonometry-aware neural networks for drug-protein binding structure
  prediction,'' in \emph{Advances in Neural Information Processing Systems}.

\bibitem{stark2022equibind}
H.~St{\"a}rk, O.~Ganea, L.~Pattanaik, R.~Barzilay, and T.~Jaakkola, ``Equibind:
  Geometric deep learning for drug binding structure prediction,'' in
  \emph{International Conference on Machine Learning}.\hskip 1em plus 0.5em
  minus 0.4em\relax PMLR, 2022, pp. 20\,503--20\,521.

\bibitem{ganea2021independent}
O.-E. Ganea, X.~Huang, C.~Bunne, Y.~Bian, R.~Barzilay, T.~Jaakkola, and
  A.~Krause, ``Independent se (3)-equivariant models for end-to-end rigid
  protein docking,'' \emph{arXiv preprint arXiv:2111.07786}, 2021.

\bibitem{yang2022protein}
C.~Yang, E.~A. Chen, and Y.~Zhang, ``Protein--ligand docking in the
  machine-learning era,'' \emph{Molecules}, vol.~27, no.~14, p. 4568, 2022.

\bibitem{masters2022deep}
M.~Masters, A.~H. Mahmoud, Y.~Wei, and M.~A. Lill, ``Deep learning model for
  flexible and efficient protein-ligand docking,'' in \emph{ICLR2022 Machine
  Learning for Drug Discovery}, 2022.

\bibitem{chatterjee2023improving}
A.~Chatterjee, R.~Walters, Z.~Shafi, O.~S. Ahmed, M.~Sebek, D.~Gysi, R.~Yu,
  T.~Eliassi-Rad, A.-L. Barab{\'a}si, and G.~Menichetti, ``Improving the
  generalizability of protein-ligand binding predictions with ai-bind,''
  \emph{Nature Communications}, vol.~14, no.~1, p. 1989, 2023.

\bibitem{trott2010autodock}
O.~Trott and A.~J. Olson, ``Autodock vina: improving the speed and accuracy of
  docking with a new scoring function, efficient optimization, and
  multithreading,'' \emph{Journal of computational chemistry}, vol.~31, no.~2,
  pp. 455--461, 2010.

\bibitem{koes2013lessons}
D.~R. Koes, M.~P. Baumgartner, and C.~J. Camacho, ``Lessons learned in
  empirical scoring with smina from the csar 2011 benchmarking exercise,''
  \emph{Journal of chemical information and modeling}, vol.~53, no.~8, pp.
  1893--1904, 2013.

\bibitem{mcnutt2021gnina}
A.~T. McNutt, P.~Francoeur, R.~Aggarwal, T.~Masuda, R.~Meli, M.~Ragoza,
  J.~Sunseri, and D.~R. Koes, ``Gnina 1.0: molecular docking with deep
  learning,'' \emph{Journal of cheminformatics}, vol.~13, no.~1, pp. 1--20,
  2021.

\bibitem{tran2015learning}
D.~Tran, L.~Bourdev, R.~Fergus, L.~Torresani, and M.~Paluri, ``Learning
  spatiotemporal features with 3d convolutional networks,'' in
  \emph{Proceedings of the IEEE international conference on computer vision},
  2015, pp. 4489--4497.

\bibitem{hassan2017protein}
N.~M. Hassan, A.~A. Alhossary, Y.~Mu, and C.-K. Kwoh, ``Protein-ligand blind
  docking using quickvina-w with inter-process spatio-temporal integration,''
  \emph{Scientific reports}, vol.~7, no.~1, p. 15451, 2017.

\bibitem{friesner2004glide}
R.~A. Friesner, J.~L. Banks, R.~B. Murphy, T.~A. Halgren, J.~J. Klicic, D.~T.
  Mainz, M.~P. Repasky, E.~H. Knoll, M.~Shelley, J.~K. Perry \emph{et~al.},
  ``Glide: a new approach for rapid, accurate docking and scoring. 1. method
  and assessment of docking accuracy,'' \emph{Journal of medicinal chemistry},
  vol.~47, no.~7, pp. 1739--1749, 2004.

\bibitem{xu2018powerful}
K.~Xu, W.~Hu, J.~Leskovec, and S.~Jegelka, ``How powerful are graph neural
  networks?'' \emph{arXiv preprint arXiv:1810.00826}, 2018.

\bibitem{jing2020learning}
B.~Jing, S.~Eismann, P.~Suriana, R.~J. Townshend, and R.~Dror, ``Learning from
  protein structure with geometric vector perceptrons,'' \emph{arXiv preprint
  arXiv:2009.01411}, 2020.

\bibitem{rogers2010extended}
D.~Rogers and M.~Hahn, ``Extended-connectivity fingerprints,'' \emph{Journal of
  chemical information and modeling}, vol.~50, no.~5, pp. 742--754, 2010.

\bibitem{liu2015pdb}
Z.~Liu, Y.~Li, L.~Han, J.~Li, J.~Liu, Z.~Zhao, W.~Nie, Y.~Liu, and R.~Wang,
  ``Pdb-wide collection of binding data: current status of the pdbbind
  database,'' \emph{Bioinformatics}, vol.~31, no.~3, pp. 405--412, 2015.

\bibitem{deng2010multiple}
H.~Deng, C.~J. Doonan, H.~Furukawa, R.~B. Ferreira, J.~Towne, C.~B. Knobler,
  B.~Wang, and O.~M. Yaghi, ``Multiple functional groups of varying ratios in
  metal-organic frameworks,'' \emph{Science}, vol. 327, no. 5967, pp. 846--850,
  2010.

\bibitem{hua2022cpinformer}
Y.~Hua, X.~Song, Z.~Feng, X.-J. Wu, J.~Kittler, and D.-J. Yu, ``Cpinformer for
  efficient and robust compound-protein interaction prediction,''
  \emph{IEEE/ACM transactions on computational biology and bioinformatics},
  vol.~20, no.~1, pp. 285--296, 2022.

\bibitem{vaswani2017attention}
A.~Vaswani, N.~Shazeer, N.~Parmar, J.~Uszkoreit, L.~Jones, A.~N. Gomez,
  {\L}.~Kaiser, and I.~Polosukhin, ``Attention is all you need,''
  \emph{Advances in neural information processing systems}, vol.~30, 2017.

\bibitem{tholke2022torchmd}
P.~Th{\"o}lke and G.~De~Fabritiis, ``Torchmd-net: equivariant transformers for
  neural network based molecular potentials,'' \emph{arXiv preprint
  arXiv:2202.02541}, 2022.

\bibitem{liao2022equiformer}
Y.-L. Liao and T.~Smidt, ``Equiformer: Equivariant graph attention transformer
  for 3d atomistic graphs,'' \emph{arXiv preprint arXiv:2206.11990}, 2022.

\bibitem{morehead2021geometric}
A.~Morehead, C.~Chen, and J.~Cheng, ``Geometric transformers for protein
  interface contact prediction,'' \emph{arXiv preprint arXiv:2110.02423}, 2021.

\bibitem{satorras2021n}
V.~G. Satorras, E.~Hoogeboom, and M.~Welling, ``E (n) equivariant graph neural
  networks,'' in \emph{International conference on machine learning}.\hskip 1em
  plus 0.5em minus 0.4em\relax PMLR, 2021, pp. 9323--9332.

\bibitem{landrum2013rdkit}
G.~Landrum \emph{et~al.}, ``Rdkit: A software suite for cheminformatics,
  computational chemistry, and predictive modeling,'' \emph{Greg Landrum},
  vol.~8, 2013.

\bibitem{krivak2018p2rank}
R.~Kriv{\'a}k and D.~Hoksza, ``P2rank: machine learning based tool for rapid
  and accurate prediction of ligand binding sites from protein structure,''
  \emph{Journal of cheminformatics}, vol.~10, pp. 1--12, 2018.

\bibitem{burley2021rcsb}
S.~K. Burley, C.~Bhikadiya, C.~Bi, S.~Bittrich, L.~Chen, G.~V. Crichlow, C.~H.
  Christie, K.~Dalenberg, L.~Di~Costanzo, J.~M. Duarte \emph{et~al.}, ``Rcsb
  protein data bank: powerful new tools for exploring 3d structures of
  biological macromolecules for basic and applied research and education in
  fundamental biology, biomedicine, biotechnology, bioengineering and energy
  sciences,'' \emph{Nucleic acids research}, vol.~49, no.~D1, pp. D437--D451,
  2021.

\bibitem{zhu2022torchdrug}
Z.~Zhu, C.~Shi, Z.~Zhang, S.~Liu, M.~Xu, X.~Yuan, Y.~Zhang, J.~Chen, H.~Cai,
  J.~Lu \emph{et~al.}, ``Torchdrug: A powerful and flexible machine learning
  platform for drug discovery,'' \emph{arXiv preprint arXiv:2202.08320}, 2022.

\end{thebibliography}
\end{document}